\documentclass{article}

\def\makinhome{/home/makin}
\makeatletter
\begingroup\endlinechar=-1\relax
    \everyeof{\noexpand}%
    \edef\x{\endgroup\def\noexpand\homepath{%
        \@@input|"kpsewhich --var-value=HOME" }}\x
\makeatother

\ifx\homepath\makinhome
    \newcommand{\stydir}{../../stys}
    \newcommand{\bibsdir}{../../bibs}
    \newcommand{\tikzdir}{../../tikzpics/ecog2vec2txt}
    \newcommand{\figdir}{../../../jpgs/ecog2vec2txt}
\else
    \newcommand{\stydir}{stys}
    \newcommand{\bibsdir}{bibs}
    \newcommand{\tikzdir}{tikzpics}
    \newcommand{\figdir}{figs}
\fi

\usepackage[preprint]{\stydir/neurips_2024}

\providecommand{\stydir}{../stys}
\providecommand{\bibsdir}{../bibs}
\providecommand{\tikzdir}{../tikzpics}

\usepackage{%
	amsmath,amsfonts,
	amssymb,amsthm,graphicx,rotating,booktabs,
	versions,ifthen,siunitx,bbm
}
\usepackage{xcolor}
\usepackage{pgf,tikz,pgfplots} 
\usetikzlibrary{%
 	positioning,calc,patterns,shapes,arrows,intersections,backgrounds,fit,shadings,decorations.text,shadows,fadings,bayesnet,
}
\usepgfplotslibrary{colorbrewer,colormaps,fillbetween,groupplots}

\ifdefined\rvmacroize
\else
	\makeatletter
	\input{\stydir/jgmstd.sty}
	\makeatother
\fi

\newboolean{INCLFIGS} 
\setboolean{INCLFIGS}{true}

\usepackage{caption,subfig}

\ifthenelse{\boolean{INCLFIGS}}{
	\captionsetup[subfloat]{font={Large,sf},labelformat=simple,labelsep=none}
}{
	\captionsetup[subfloat]{labelformat=empty,labelsep=none}
}

\newcommand{\colorprovide}[2]{\@ifundefinedcolor{#1}{\colorlet{#1}{#2}}{}}

\usepackage{multirow}
\usepackage{hyperref}
\usepackage{url}
\usepackage{wrapfig}


\newcommand{\intercolarrow}{%
    \begin{tikzpicture}
        \node at (0,0) {};
        \draw[->] (0,0.5) -- (0.6,0.5);
    \end{tikzpicture}%
}

\newcommand{\TablePerformanceVsTrainingSize}{%
\begin{table}[ht]
    \centering
    \caption{Median percentage improvement in WER provided by context vectors over the original ECoG data, for various numbers of training blocks ($N_\text{blocks}$) for the \ecogtotxt\ model.
    Stars indicate that the distribution of improvements is significantly above zero (Wilcoxon signed-rank test; * $p<0.05$; ** $p<0.005$). Exact p-values can be found in \tbl{cross_validation_p}.}\vspace{\baselineskip}
    \begin{tabular}{@{}cccccccc@{}}
        \centering
        \begin{tabular}{@{}clclclcl@{}}
            \toprule
            \multicolumn{2}{c}{\parta} & \multicolumn{2}{c}{\partb} & \multicolumn{2}{c}{\partc} & \multicolumn{2}{c}{\partd}\\
            \midrule
            $N_\text{blocks}$ & \% improve & $N_\text{blocks}$ & \% improve & $N_\text{blocks}$ & \% improve & $N_\text{blocks}$ & \% improve \\
            \midrule
            1 & 2.56\%  & 1 & 21.91\%** & 4  & 7.16\%**  & 4  & 2.93\%      \\
            2 & 0.17\%  & 2 & 45.21\%** & 7  & 20.91\%** & 9  & 20.47\%** \\
              &         & 3 & 59.90\%** & 10 & 23.58\%** & 15 & 20.82\%** \\
              &         & 4 & 55.81\%** & 13 & 21.42\%** & 20 & 14.72\%   \\
              &         & 5 & 56.15\%** & 16 & 15.85\%*  & 26 & 9.95\%    \\
              &         & 6 & 46.36\%** & 19 & 28.38\%** & 31 & 9.43\%    \\
              &         & 7 & 25.38\%   & 22 & 35.32\%*  & 37 & -5.52\%   \\
              &         & 8 & 34.95\%*  & 25 & 28.68\%*  & 42 & -9.94\%   \\
              &         & 9 & 50.19\%*  &    &           &    &           \\
            \bottomrule
        \end{tabular}
    \end{tabular}
    \label{tbl:performanceVsTrainingSize}
\end{table}
}

\newcommand{\TableWVlossVsETloss}{%
\begin{wraptable}{r}{2in}
    \caption{\wavtovec\ losses and accuacies by participant.}
    \begin{tabular}{@{}ccc@{}}
        \toprule
        Participant & Best loss & Accuracy \\
        \midrule
        \parta & 0.451  & 0.845 \\
        \partb & 1.159  & 0.711 \\
        \partc & 1.574  & 0.600 \\
        \partd & 1.321  & 0.667 \\
        \bottomrule
    \end{tabular}
    \label{tbl:WVlossVsETloss}
\end{wraptable}
}

\newcommand{\TablePvals}{
    \begin{table}[ht]
        \centering
        \caption{For each participant, p-values for the null hypothesis that the distribution of WERs is not lower (better) for context vectors than ECoG data (one-sided Wilcoxon signed-rank test).}
        \label{tbl:cross_validation_p}
        \begin{tabular}{@{}cccc@{}}
            \toprule
            participant & no. training & test statistic & p-value \\
             & blocks &  &  \\
            \midrule
            \multirow{2}{*}{\parta}
            & 1 & 240.0 & 5.64e-1 \\
            & 2 & 206.0 & 2.99e-1 \\
            \midrule
            \multirow{9}{*}{\partb}
            & 1 & 73.0 & 3.04e-4 \\
            & 2 & 1.0 & 1.86e-9 \\
            & 3 & 24.0 & 7.10e-7 \\
            & 4 & 42.0 & 1.04e-5 \\
            & 5 & 92.0 & 1.49e-3 \\
            & 6 & 88.0 & 1.09e-3 \\
            & 7 & 156.0 & 5.95e-2 \\
            & 8 & 132.0 & 1.92e-2 \\
            & 9 & 123.0 & 1.17e-2 \\
            \midrule
            \multirow{8}{*}{\partc}
            & 4 & 85.0 & 8.59e-4 \\
            & 7 & 79.0 & 5.19e-4 \\
            & 10 & 84.0 & 7.92e-4 \\
            & 13 & 103.0 & 3.32e-3 \\
            & 16 & 133.0 & 2.02e-2 \\
            & 19 & 105.0 & 3.81e-3 \\
            & 22 & 129.0 & 1.64e-2 \\
            & 25 & 131.0 & 1.82e-2 \\
            \midrule
            \multirow{8}{*}{\partd}
            & 4 & 165.0 & 8.53e-2 \\
            & 9 & 16.0 & 1.57e-7 \\
            & 15 & 51.0 & 3.14e-5 \\
            & 20 & 160.0 & 7.01e-2 \\
            & 26 & 157.0 & 6.21e-2 \\
            & 31 & 179.0 & 1.40e-1 \\
            & 37 & 247.0 & 6.19e-1 \\
            & 42 & 270.0 & 7.80e-1 \\
            \bottomrule
        \end{tabular}
    \end{table}
}

\newcommand{\TableMochaPvals}{
    \begin{table}[ht]
        \centering
        \caption{For participants \parta, \partb, and \partd, p-values for the null hypothesis that the distribution of WERs is not lower (better) for context vectors than ECoG data (Wilcoxon signed-rank test).}
        \label{tbl:cross_validation_p_mocha}
        \begin{tabular}{@{}cccc@{}}
            \toprule
            participant & $H_{a}$ & test statistic & p-value \\
            \midrule
            \multirow{2}{*}{\parta}
            & context $<$ ECoG-ID & 206.0 & 2.99e-1 \\
            & context $\neq$ ECoG-all & 0.0 & 1.86e-9 \\
            \midrule
            \multirow{2}{*}{\partb}
            & context $<$ ECoG-ID & 1.0 & 1.86e-9 \\
            & context $\neq$ ECoG-all & 187.0 & 3.60e-1 \\
            \midrule
            \multirow{2}{*}{\partd}
            & context $<$ ECoG-ID & 24.0 & 7.10e-7 \\
            & context $\neq$ ECoG-all & 70.0 & 4.60e-4 \\
            \bottomrule
        \end{tabular}
    \end{table}
}

\newcommand{\TableEffectOfLookahead}{%
    \begin{wraptable}{r}{3in}
        \centering
        \caption{The effect of $s$, the number of steps into the future that \wavtovec\ must predict, on (final) \wavtovec\ loss and accuracy, and speech-decoding word error rates (means and variances) for participant \partc's ECoG recordings.}
        \label{tbl:effectOfLookahead}
        \small
        \begin{tabular}{@{}ccccc@{}}
            \toprule
            $s$ & Best Loss & Accuracy & $\bar{x}_{\textnormal{WER}}$ & $s^2_{\textnormal{WER}}$ \\
            \midrule
            2  & 1.246 & 0.689 & 0.240 & 0.025 \\
            4  & 1.433 & 0.617 & 0.230 & 0.026 \\
            6  & 1.511 & 0.594 & 0.220 & 0.026 \\
            8  & 1.574 & 0.596 & 0.205 & 0.026 \\
            10 & 1.634 & 0.580 & 0.218 & 0.027 \\
            12 & 1.687 & 0.355 & 0.227 & 0.027 \\
            14 & 1.747 & 0.548 & 0.214 & 0.029 \\
            16 & 1.788 & 0.532 & 0.219 & 0.026 \\
            18 & 1.831 & 0.520 & 0.219 & 0.028 \\
            20 & 1.890 & 0.506 & 0.225 & 0.026 \\
            \bottomrule
        \end{tabular}
    \end{wraptable}
}

\title{Improving Speech Decoding from ECoG with Self-Supervised Pretraining}

\author{%
  Brian A.~Yuan\\
  Institute of Neuroinformatics\\
  University of Zurich and ETH Zurich\\
  Zurich, Switzerland \\
  \texttt{bryuan@student.ethz.ch} \\
  \And
  Joseph G.~Makin \\
  School of Electrical and Computer Engineering \\
  Purdue University \\
  West Lafayette, IN, USA \\
  \texttt{jgmakin@purdue.edu} \\
}

\usetikzlibrary{pgfplots.statistics}

\newcommand{\wavtovec}{{\sc wav2vec}}
\newcommand{\Wavtovec}{{\sc Wav2vec}}
\newcommand{\ecogtotxt}{{\sc ecog2txt}}

\newcommand{\parta}{{\bfseries a}}
\newcommand{\partb}{{\bfseries b}}
\newcommand{\partc}{{\bfseries c}}
\newcommand{\partd}{{\bfseries d}}

\rvmacroize[!]{ltnt}

\rvmacroize[!]{context}

\newcommand{\ie}{\emph{i.e.}}

\begin{document}

\maketitle

\begin{abstract}
    Recent work on intracranial brain-machine interfaces has demonstrated that spoken speech can be decoded with high accuracy, essentially by treating the problem as an instance of supervised learning and training deep neural networks to map from neural activity to text. However, such networks pay for their expressiveness with very large numbers of labeled data, a requirement that is particularly burdensome for invasive neural recordings acquired from human patients. On the other hand, these patients typically produce speech outside of the experimental blocks used for training decoders. Making use of such data, and data from other patients, to improve decoding would ease the burden of data collection---especially onerous for dys- and anarthric patients. Here we demonstrate that this is possible, by reengineering {\sc{wav2vec}}---a simple, self-supervised, fully convolutional model that learns latent representations of audio using a noise-contrastive loss---for electrocorticographic (ECoG) data. We train this model on unlabelled ECoG recordings, and subsequently use it to transform ECoG from labeled speech sessions into {\sc{wav2vec}}'s representation space, before finally training a supervised encoder-decoder to map these representations to text. We experiment with various numbers of labeled blocks; for almost all choices, the new representations yield superior decoding performance to the original ECoG data, and in no cases do they yield worse. Performance can also be improved in some cases by pretraining {\sc{wav2vec}} on another patient's data. In the best cases, {\sc{wav2vec}}'s representations decrease word error rates over the original data by upwards of 50\%.
\end{abstract}

\section{Introduction}\label{sec:intro}
In the last five years, decoding speech from neural data has moved from sub-chronic studies with epilepsy and tumor patients 
\cite{Pei2011,Mugler2018,Angrick2019,Moses2019,Anumanchipalli2019,Sun2020,Makin2020}
into clinical trials with persons who have lost the ability to speak
\cite{Moses2021,Metzger2023,Willett2023}.
However, word error rates (WERs) on vocabularies of $\sim$1000 words or higher are typically around 25\%---at the outer edge of usability \cite{Munteanu06} and not clearly superior to non-invasive alternatives built around eye or head tracking.
Furthermore, for electrocorticography (ECoG), achieving such WERs requires $\sim$15--20 hours of data, which in practice with non-speaking patients takes about two weeks to collect \cite{Metzger2023}.
When decoding from single units like the Utah array \cite{Willett2023}, far fewer data are needed---about 200 sentences, or approximately one hour of data collection; but decoding noticeably deteriorates over just one week, so the decoder must be regularly recalibrated.

On the other hand, with an implanted array, 24 hours of data can conceivably be recorded every day, even though the vast majority of this time lies outside of controlled sessions.
Making use of these data to improve decoding, even with much lower
``bang for the buck,''
would be ``free money,''
since at present they are entirely wasted.

Happily, the past half decade of machine-learning research has also seen the development and successful application of a suite of unsupervised and ``self-supervised'' techniques for learning representations from input data alone.
For example, a neural network can be trained on sequence data to aggregate representations that are highly predictive of future elements of the sequence, or transformations thereof \cite{vandenOord2018,Schneider2019,Baevski2020}.
When applied to raw, unlabelled speech audio, the aggregated representations also make very good features for assigning phonemes or characters to the audio sequence, that is, for underwriting automatic speech recognition.
Indeed, accuracies competitive with fully supervised methods can be achieved merely by ``fine-tuning'' the network---or even a simple linear, final layer---on a much smaller set of phoneme-labeled audio \cite{vandenOord2018,Schneider2019,Baevski2020}.
Such techniques have recently been shown to be effective on neural data \cite{Schneider2023}.

In this study, we demonstrate the effectiveness of such techniques for improving the decoding of speech from sub-chronic electrocorticograms (ECoG).
In particular, we revisit the results of Makin and colleagues \cite{Makin2020}, in which participants undergoing monitoring for epilepsy read aloud sentences from small data sets (30--50 sentences, 125--250 words).
An encoder-decoder neural network was then trained to map the electrocorticogram recorded during each spoken sentence to the
corresponding sequences of words, one word at a time.
Performance was evaluated by calculating the encoder-decoder's word error rates on a block held out from training, albeit one consisting of those same 30--50 sentences.

However, these participants also typically read aloud $\sim$400 other unique sentences, which did not form part of the training or testing sets.
We refer to these sentences and the corresponding ECoG as ``out-of-distribution'' (OOD) data.
Here we use the ECoG data recorded during these sessions---but not labels of any kind---to train instances of \wavtovec, a deep convolutional neural network originally designed for unsupervised learning of useful representations from speech audio.
We then use the trained network to ``aggregate'' \emph{in-distribution} ECoG data into what we anticipate will be more useful representations for the downstream task of speech decoding.
Finally, we train encoder-decoder networks to map the aggregated in-distribution ECoG data to text.
We find that for most participants and most sizes of data for the supervised-learning task, the aggregated representations produced by \wavtovec\ yield superior speech decoding to that of the original ECoG data.

\section{Methods}\label{sec:methods}
We summarize the entire pipeline for transforming raw ECoG into text in \fig{schematic}.
Below we provide the details.
All code is available at
\url{https://github.com/b4yuan/ecog2vec}
under the MIT license.

\subsection{Data}\label{sec:data}
We consider a dataset (provided upon request by the original authors \cite{Makin2020}) in which participants undergoing treatment for epilepsy read aloud sentences while neural activity was recorded with an intracranial grid of electrodes (ECoG).
Each of several recording ``blocks'' of sentences consisted of either a subset of 460 sentences/$\sim$1800 unique words (MOCHA-TIMIT \cite{Wrench1999} subsets), or a set of 30 sentences/$\sim$125 unique words (``picture descriptions'').
Following Makin and colleagues \cite{Makin2020}, for each participant, we trained and tested speech decoders only on the first 50-sentence ($\sim$150 unique words) subset of MOCHA-TIMIT (henceforth ``MOCHA-1'') or the picture descriptions---whichever set the participant had read more of (MOCHA-1 for participants \parta\ and \partb; picture descriptions for participants \partc\ and \partd).
For brevity, we refer to these as ``in-distribution'' (ID) data.
The remaining (``out-of-distribution,'' OOD) sentences were used without labels for self-supervised pretraining (see \sctn{selfsupervisedmethods} below).
Participants \parta, \partc, and \partd\ had roughly an hour of OOD data; \partb\ had roughly 30 minutes.

Simultaneous with speech, electrocorticograms were recorded from study participants from the cortical areas near the Sylvian fissure, including areas associated with speech production and perception (see the original study for details \cite{Makin2020}).
Electrode grids consisted of either 256 (\parta, \partb, \partd) or 128 (\partc) nominal electrodes per participant, with 4-mm spacing.
The analog ECoG signals were digitized at roughly 3 kHz, 
and channels
with visible artifacts or noise were removed.
The remaining electrodes were then notch filtered at 60 Hz to remove line noise and then band-pass filtered into the ``high-$\gamma$'' range, 70--200 Hz.
The envelopes (analytic amplitudes) of these signal were then computed.
Finally, all channels were re-referenced against their common average (CAR).
In short, our preprocessing pipeline followed the original protocol except that we used a CAR in place of bipolar referencing and set our filter cut-off at 200 rather than 150 Hz.

\subsection{Self-supervised training}\label{sec:selfsupervisedmethods}

\paragraph{Data preparation.} 
For each participant, all ECoG data (at $\sim$3 kHz) for all OOD blocks (\ie, excluding the MOCHA-1 blocks for participants \parta\ and \partb\ and picture description blocks for participants \partc\ and \partd) were split into 200,000-sample long segments without removing silent periods between spoken sentences.
These segments---roughly one minute long---were then split randomly into 90\% training and 10\% validation sets for \wavtovec.

\paragraph{Architecture and hyperparameters.} 
\Wavtovec\footnote{\url{https://github.com/facebookresearch/fairseq/tree/main/examples/wav2vec}; released under the MIT license.} is a deep convolutional neural network designed to extract useful representations from a single channel of audio waveform \cite{Schneider2019}.
This is accomplished by training the network to produce activities in its deepest layer (``context vectors,'' $\contexts{t}$) that are highly informative about activities at an intermediate layer at future points in time ($\ltnts{t+s}$).
More precisely, the network is trained to discriminate latent vectors $\ltnts{t+s}$ for which $\contexts{t}$ provides context, from latent vectors from other contexts.
It can be shown that this objective forces the network to maximize (a lower bound on) mutual information between $\contexts{t}$ that $\ltnts{t+s}$ \cite{vandenOord2018}.

To adapt \wavtovec\ to ECoG data, we expand the initial convolutional layer to have $N_\text{electrodes}$ input channels.
This varies by participant and is between 128 and 245 (since noisy channels are removed first; see above).
Since our dataset is about three orders of magnitude smaller than a typical audio dataset (Librispeech, $\sim$1000 hours vs.\ $\sim$1 hour per participant), we also need to scale the model down significantly.
We use three convolutional layers between the ECoG data and the latent variable $\ltnts{}$, with kernel sizes (8, 4, 2) and strides (5, 3, 2).
This brings the sampling rate down to $\sim$100 Hz, which we chose because (1) it approximately matches the order of the phoneme-production rate (20--30 phonemes/second), and (2) it was the effective sampling rate of the latent vectors in the original \wavtovec\ used for speech audio.
We then use three additional convolutional layers between $\ltnts{}$ and the output (context) vectors $\contexts{}$, with kernel sizes (2, 3, 4) and a stride of 1.
This yields a receptive field at the final layer of about 70 ms. 
All convolutions are causal and have 512 output channels.
In contrast to the original implementation, we use ELU activation functions throughout, which improved performance during some experiments.

\TableEffectOfLookahead

The time difference, $s$, between context vectors $\contexts{t}$ and latent states $\ltnts{t+s}$ which are to be maximally mutually informative is a hyperparameter and must be selected ``by hand.''
We identified this value, $s=8$, by examining speech-decoding results for a single participant (\partc), which are shown in \tbl{effectOfLookahead}.
This allows a small amount of information to leak from the test set into the training set for participant \partc; we emphasize that we kept $s=8$ for all other participants.

\paragraph{Training.}
All \wavtovec\ models were trained by stochastic gradient descent of the \wavtovec\ loss \cite{Schneider2019} (a variant on the {\sc InfoNCE} loss \cite{vandenOord2018}) with the AdaM optimizer and a learning rate of $1\mathrm{e}{-4}$.
Training was continued until the validation loss did not improve over fifty consecutive epochs (``convergence'').

\paragraph{Transfer learning.}
For all participants, we also explored a form of transfer learning in which \wavtovec\ was first trained on one participant's OOD data and subsequently ``fine-tuned'' on another participant's OOD data.
More precisely, we first trained an instance of \wavtovec\ on one participant's data until convergence.
Because number and location of electrodes varied by participant, each requires at least one ``proprietary'' layer in \wavtovec.
Accordingly, we removed the initial convolutional layer from the trained model and replaced it with a randomly initialized layer with number of input channels matched to the second (target) participant.
We then froze weight updates for all layers except this input layer, and trained for 100 epochs on the second participant's data.
Finally, we unfroze the entire network and trained until the validation loss did not improve for fifty consecutive epochs.

    \begin{figure}[ht]
        \centering%
        \footnotesize%
        \begin{tikzpicture}[node distance=1cm]
            \node (preprocessing) at (-5,0){%
                \setlength\tabcolsep{1pt}%
                \begin{tabular}{@{}cccccccccc@{}}%
                    \includegraphics[width=2cm]{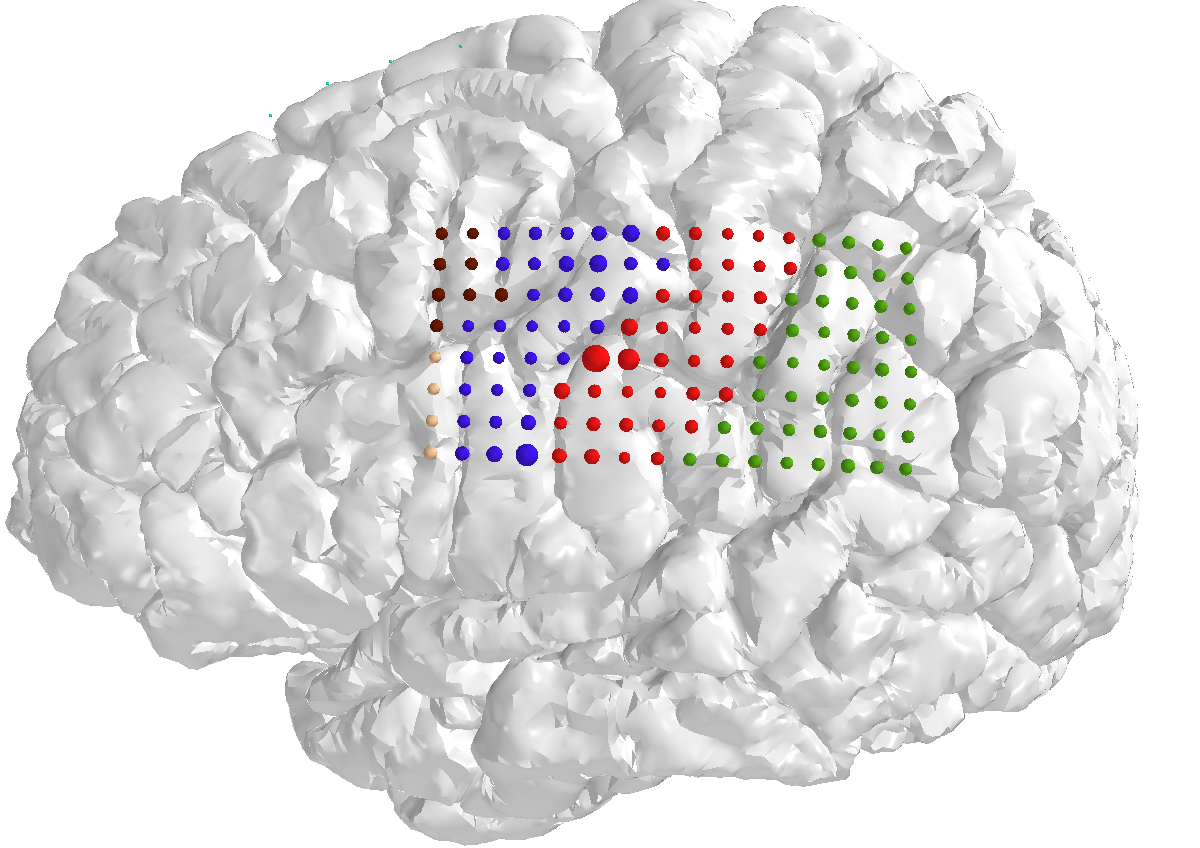} &
                    \intercolarrow
                    &
                    \providecommand{\figheight}{1.6in}%
                    \providecommand{\figwidth}{2.2in}%
                    \hspace{8px}
                    \input{\tikzdir/e2v2t_schematic/notch} &
                    \intercolarrow
                    &
                    \providecommand{\figheight}{2.2in}%
                    \providecommand{\figwidth}{2.2in}%
\begin{tikzpicture}

\definecolor{darkgray176}{RGB}{176,176,176}%
\definecolor{darkorange25512714}{RGB}{255,127,14}%
\definecolor{forestgreen4416044}{RGB}{44,160,44}%
\definecolor{lightgray204}{RGB}{204,204,204}%
\definecolor{steelblue31119180}{RGB}{31,119,180}%
\providecommand{\figwidth}{2in}
\providecommand{\figheight}{2in}

\begin{axis}[
legend cell align={left},
legend style={
  fill opacity=0.8,
  draw opacity=1,
  text opacity=1,
  at={(0.03,0.97)},
  anchor=north west,
  draw=lightgray204
},
tick align=outside,
tick pos=left,
x grid style={darkgray176},
xmin=-10, xmax=210,
y grid style={darkgray176},
ymin=-28.9576784780365, ymax=10,
ytick style={color=black},
scale=0.3,
ytick={}, 
yticklabels={}, 
xtick=\empty,
ytick=\empty,
axis lines=none,
width=\figwidth,
height=\figheight
]
\addplot [semithick, steelblue31119180]
table {%
0 -7.17369857738959
1 -7.20467187420581
2 -6.13826978224097
3 -3.41740292242321
4 0.915553641789302
5 5.58856845600531
6 6.2273938965518
7 1.84617874765536
8 -0.214484614957655
9 1.3935326705905
10 2.44623788603349
11 2.03497847905965
12 2.66666302195517
13 6.45531463305815
14 4.07225070375716
15 0.546001331258594
16 -1.06095615137747
17 -3.41209943144349
18 -3.28794794768328
19 -4.24869767812197
20 -6.17443129158346
21 -9.13210533326492
22 -12.2323308460182
23 -11.0994751594262
24 -12.5195347209228
25 -14.4096775329672
26 -13.0450616779854
27 -14.6254833452986
28 -15.7314116222551
29 -17.498114175396
30 -20.2886840270367
31 -20.9869431273546
32 -19.1431663552066
33 -16.101306755445
34 -16.4267912623473
35 -17.9941380338278
36 -17.0422044902807
37 -15.1429330799147
38 -18.9449929166585
39 -19.0570954146096
40 -16.1824282258749
41 -17.3816370079294
42 -15.6825026351726
43 -16.4279990713112
44 -16.4729863172397
45 -14.8186609294498
46 -13.348913853406
47 -10.3316369859385
48 -11.0205874079838
49 -11.343373444106
50 -9.44467592489673
51 -9.58036980591714
52 -6.94371055942611
53 -4.03147078031907
54 -2.7305052299198
55 1.93327582564962
56 6.05861396252294
57 5.35804520040983
58 0.0537956168500386
59 -1.63184051871212
60 1.83702866252133
61 1.61776517870749
62 0.323503172694473
63 3.25982523463608
64 4.4387729758455
65 1.65316782840819
66 -0.403458642495025
67 -4.370365786599
68 -6.04873093834613
69 -7.07388926457497
70 -9.26694610825507
71 -10.8904714579694
72 -14.3850693348213
73 -16.9173199537909
74 -16.3507884280989
75 -18.1712239282206
76 -18.8561280083377
77 -18.8792837434448
78 -19.8728484974708
79 -20.1680450118147
80 -23.1433823500993
81 -25.2869413088774
82 -25.7814244832844
83 -25.0175999099156
84 -21.6286680370104
85 -22.0248984987848
86 -23.6203286476666
87 -23.7584190472262
88 -22.6487518375507
89 -23.321224944084
90 -24.3682407017332
91 -22.4195573537145
92 -19.7623485291842
93 -18.069955331157
94 -17.7566107595339
95 -17.7087495103478
96 -16.5888195624575
97 -13.3473204186885
98 -10.1114892459009
99 -10.5278022601851
100 -10.3492711787112
101 -8.4917828644393
102 -8.75449313753052
103 -7.48489128454821
104 -6.00609564571641
105 -4.98620693178964
106 -0.997327219920408
107 3.20225967698207
108 0.830754856906424
109 -4.98910458190949
110 -4.71885732622468
111 -1.64402979407896
112 -0.680668961194897
113 -0.795011942500423
114 1.80177562469908
115 5.59686304768547
116 3.81286781703238
117 0.826681571197696
118 -0.330853964669586
119 -0.536599827682949
120 -1.22315111639182
121 -1.4386832845048
122 -1.82189580755221
123 -5.34344644620433
124 -7.02742454450345
125 -5.8033529057866
126 -7.89974910730962
127 -8.12051075627096
128 -8.10320852906443
129 -9.91928482108051
130 -10.0172492238926
131 -12.9536983877188
132 -15.3594046423677
133 -15.9716109919827
134 -14.2139460876933
135 -10.4771570477169
136 -13.1215847432031
137 -15.9346600412391
138 -15.2191942106583
139 -16.7502548720222
140 -20.0281920115231
141 -20.6757867999841
142 -19.780683942372
143 -18.8393441931112
144 -18.5500593943289
145 -19.4137101061642
146 -19.1449234989705
147 -17.0490602613427
148 -13.3948406073614
149 -11.0609089460922
150 -11.8647576528019
151 -11.1855715658749
152 -9.89271029538941
153 -8.45281556394184
154 -6.190740805323
155 -4.88370142193162
156 -1.89307695563912
157 2.45632418227615
158 6.19480260866112
159 4.46576268586796
160 -1.39143946853437
161 0.151950303006743
162 2.25986423174618
163 1.28526073694957
164 1.45047840760526
165 5.09331584908068
166 7.83797895564931
167 5.29518683833885
168 2.77638582701911
169 -0.146770972264676
170 -0.771661632370524
171 0.0634525108011985
172 -0.611106713677145
173 -3.16411615131074
174 -6.86672274241573
175 -9.15398959477898
176 -9.91540855466155
177 -12.0322829388897
178 -12.5088445201982
179 -13.0722701214836
180 -14.2776161737856
181 -14.5827443702728
182 -17.6466601260472
183 -18.8021003850736
184 -17.2312866197899
185 -15.0431369547732
186 -13.1258057081141
187 -13.9765643325518
188 -14.6901757034357
189 -12.3872605399811
190 -11.9196656669374
191 -16.0569161380408
192 -15.988767700037
193 -14.6774937093141
194 -12.7841194625944
195 -10.571943676041
196 -11.3811856863322
197 -11.4568010758376
198 -9.05298384168418
199 -5.29583167008241
};
\addplot [semithick, darkorange25512714]
table {%
0 -0.604918613510914
1 -0.461676507422694
2 -0.24612876690563
3 0.0391534442357597
4 0.383388869723101
5 0.768224718825181
6 1.16928789970089
7 1.55842495116537
8 1.9063353789086
9 2.18528455399463
10 2.37160750601259
11 2.44776891659372
12 2.40381330854378
13 2.23811309296168
14 1.95739514080712
15 1.57609505508698
16 1.11514796155485
17 0.600369513749562
18 0.060604654335274
19 -0.474179408805962
20 -0.97470463388195
21 -1.41398946773005
22 -1.76869436142506
23 -2.02019566780293
24 -2.15541327656701
25 -2.16743260818739
26 -2.05595372340142
27 -1.82757138348704
28 -1.49584796374585
29 -1.08109930865202
30 -0.609787426979086
31 -0.113417154397518
32 0.373125876474878
33 0.813783606756618
34 1.17397087445313
35 1.42344979413546
36 1.53914788295098
37 1.50759654332275
38 1.3266773732269
39 1.0064232885215
40 0.568720616413513
41 0.0458821022782635
42 -0.521808623117497
43 -1.08936056744494
44 -1.61083190346433
45 -2.04320691117815
46 -2.35000665992924
47 -2.50427746644687
48 -2.49067244942971
49 -2.30644447642689
50 -1.9612914314856
51 -1.47612315460125
52 -0.880938597089397
53 -0.212096119818363
54 0.490685188091767
55 1.18725056591612
56 1.83956730696258
57 2.41371706604185
58 2.88124891775529
59 3.21994520376956
60 3.41414963065308
61 3.45485171607049
62 3.33970345624823
63 3.07307133370849
64 2.66612369432156
65 2.13685230296989
66 1.50985803335308
67 0.815714501623575
68 0.0897653068503969
69 -0.629699481394644
70 -1.30382474388953
71 -1.89594807536838
72 -2.37445670268506
73 -2.71541295080968
74 -2.90461714176223
75 -2.93882998418371
76 -2.82597427097738
77 -2.58426119251361
78 -2.24031908409795
79 -1.82652196785886
80 -1.37780633616177
81 -0.928318324974294
82 -0.50824737331885
83 -0.141179106966862
84 0.157755758908842
85 0.382743083743273
86 0.53706933386733
87 0.631859110977417
88 0.683919278913145
89 0.712952510324997
90 0.738484770281449
91 0.776891862764515
92 0.838902344388156
93 0.927897873205679
94 1.03923476799011
95 1.16068519783142
96 1.27395886085964
97 1.357132407545
98 1.38769917952449
99 1.34586900378842
100 1.21770665635317
101 0.997703519479059
102 0.690429782674548
103 0.311009113255406
104 -0.115714762908783
105 -0.557300824374602
106 -0.976946014961732
107 -1.33750431155403
108 -1.60583275732603
109 -1.75692272446396
110 -1.7772528452054
111 -1.66685693620529
112 -1.43974075300091
113 -1.12249392500366
114 -0.751198649184378
115 -0.366992894747178
116 -0.0108582004660797
117 0.281667045397262
118 0.485121236326805
119 0.587453311802336
120 0.591583455295323
121 0.514955726929683
122 0.387115438251674
123 0.245605262062609
124 0.130695403629795
125 0.079617656367451
126 0.121042826156925
127 0.270518761613386
128 0.527476340013908
129 0.874227423839707
130 1.27714352636974
131 1.68994271516291
132 2.0587532634559
133 2.32839409257426
134 2.44914111604818
135 2.38315888430406
136 2.10978586354741
137 1.62897619813807
138 0.962413055143317
139 0.1520941664557
140 -0.74348950337176
141 -1.6551516884391
142 -2.510101267937
143 -3.23947841709707
144 -3.78543888782015
145 -4.10695381464893
146 -4.18366080153795
147 -4.01734879032648
148 -3.63095107255558
149 -3.06522070199262
150 -2.37353416467515
151 -1.6154814948138
152 -0.850032194025606
153 -0.129105041646804
154 0.507685347559549
155 1.03647527642839
156 1.45037993100291
157 1.75836405194389
158 1.98191992508164
159 2.15001654636932
160 2.29303995573086
161 2.43661667031515
162 2.59626412055404
163 2.77372311975622
164 2.95560008155561
165 3.11460968191848
166 3.21331291482255
167 3.20985446967198
168 3.06488065564699
169 2.74861645371893
170 2.24702878023091
171 1.56610952548402
172 0.733560323504616
173 -0.202483488264905
174 -1.17763828231205
175 -2.11790010717667
176 -2.94789036492338
177 -3.59964590234766
178 -4.02077866200788
179 -4.18089178595179
180 -4.07535725467226
181 -3.72590608478887
182 -3.17790960462623
183 -2.49468186444161
184 -1.74954712077158
185 -1.0167340751119
186 -0.362332238223189
187 0.163454660422632
188 0.531703801586966
189 0.737101166394163
190 0.796934712153964
191 0.746975219046367
192 0.634922098142415
193 0.512439523672607
194 0.426994816923434
195 0.414708457478572
196 0.49523767257802
197 0.669377634129153
198 0.91963302652346
199 1.21355763140512
};
\end{axis}

\end{tikzpicture} &
                    \intercolarrow
                    &
                    \providecommand{\figheight}{3.0in}%
                    \providecommand{\figwidth}{3.0in}%
\begin{tikzpicture}

\definecolor{darkgray176}{RGB}{176,176,176}%
\definecolor{darkorange25512714}{RGB}{255,127,14}%
\definecolor{forestgreen4416044}{RGB}{44,160,44}%
\definecolor{lightgray204}{RGB}{204,204,204}%
\definecolor{steelblue31119180}{RGB}{31,119,180}%
\providecommand{\figheight}{2in}
\providecommand{\figwidth}{2in}

\begin{axis}[
legend cell align={left},
legend style={
  fill opacity=0.8,
  draw opacity=1,
  text opacity=1,
  at={(0.03,0.97)},
  anchor=north west,
  draw=lightgray204
},
tick align=outside,
tick pos=left,
x grid style={darkgray176},
xmin=-10, xmax=210,
xtick style={color=black},
y grid style={darkgray176},
ymin=-10, ymax=10,
ytick style={color=black},
scale=0.3,
xtick=\empty,
ytick=\empty,
axis lines=none,
width=\figwidth,
height=\figheight
]
\addplot [semithick, darkorange25512714]
table {%
0 -0.604918613510914
1 -0.461676507422694
2 -0.24612876690563
3 0.0391534442357597
4 0.383388869723101
5 0.768224718825181
6 1.16928789970089
7 1.55842495116537
8 1.9063353789086
9 2.18528455399463
10 2.37160750601259
11 2.44776891659372
12 2.40381330854378
13 2.23811309296168
14 1.95739514080712
15 1.57609505508698
16 1.11514796155485
17 0.600369513749562
18 0.060604654335274
19 -0.474179408805962
20 -0.97470463388195
21 -1.41398946773005
22 -1.76869436142506
23 -2.02019566780293
24 -2.15541327656701
25 -2.16743260818739
26 -2.05595372340142
27 -1.82757138348704
28 -1.49584796374585
29 -1.08109930865202
30 -0.609787426979086
31 -0.113417154397518
32 0.373125876474878
33 0.813783606756618
34 1.17397087445313
35 1.42344979413546
36 1.53914788295098
37 1.50759654332275
38 1.3266773732269
39 1.0064232885215
40 0.568720616413513
41 0.0458821022782635
42 -0.521808623117497
43 -1.08936056744494
44 -1.61083190346433
45 -2.04320691117815
46 -2.35000665992924
47 -2.50427746644687
48 -2.49067244942971
49 -2.30644447642689
50 -1.9612914314856
51 -1.47612315460125
52 -0.880938597089397
53 -0.212096119818363
54 0.490685188091767
55 1.18725056591612
56 1.83956730696258
57 2.41371706604185
58 2.88124891775529
59 3.21994520376956
60 3.41414963065308
61 3.45485171607049
62 3.33970345624823
63 3.07307133370849
64 2.66612369432156
65 2.13685230296989
66 1.50985803335308
67 0.815714501623575
68 0.0897653068503969
69 -0.629699481394644
70 -1.30382474388953
71 -1.89594807536838
72 -2.37445670268506
73 -2.71541295080968
74 -2.90461714176223
75 -2.93882998418371
76 -2.82597427097738
77 -2.58426119251361
78 -2.24031908409795
79 -1.82652196785886
80 -1.37780633616177
81 -0.928318324974294
82 -0.50824737331885
83 -0.141179106966862
84 0.157755758908842
85 0.382743083743273
86 0.53706933386733
87 0.631859110977417
88 0.683919278913145
89 0.712952510324997
90 0.738484770281449
91 0.776891862764515
92 0.838902344388156
93 0.927897873205679
94 1.03923476799011
95 1.16068519783142
96 1.27395886085964
97 1.357132407545
98 1.38769917952449
99 1.34586900378842
100 1.21770665635317
101 0.997703519479059
102 0.690429782674548
103 0.311009113255406
104 -0.115714762908783
105 -0.557300824374602
106 -0.976946014961732
107 -1.33750431155403
108 -1.60583275732603
109 -1.75692272446396
110 -1.7772528452054
111 -1.66685693620529
112 -1.43974075300091
113 -1.12249392500366
114 -0.751198649184378
115 -0.366992894747178
116 -0.0108582004660797
117 0.281667045397262
118 0.485121236326805
119 0.587453311802336
120 0.591583455295323
121 0.514955726929683
122 0.387115438251674
123 0.245605262062609
124 0.130695403629795
125 0.079617656367451
126 0.121042826156925
127 0.270518761613386
128 0.527476340013908
129 0.874227423839707
130 1.27714352636974
131 1.68994271516291
132 2.0587532634559
133 2.32839409257426
134 2.44914111604818
135 2.38315888430406
136 2.10978586354741
137 1.62897619813807
138 0.962413055143317
139 0.1520941664557
140 -0.74348950337176
141 -1.6551516884391
142 -2.510101267937
143 -3.23947841709707
144 -3.78543888782015
145 -4.10695381464893
146 -4.18366080153795
147 -4.01734879032648
148 -3.63095107255558
149 -3.06522070199262
150 -2.37353416467515
151 -1.6154814948138
152 -0.850032194025606
153 -0.129105041646804
154 0.507685347559549
155 1.03647527642839
156 1.45037993100291
157 1.75836405194389
158 1.98191992508164
159 2.15001654636932
160 2.29303995573086
161 2.43661667031515
162 2.59626412055404
163 2.77372311975622
164 2.95560008155561
165 3.11460968191848
166 3.21331291482255
167 3.20985446967198
168 3.06488065564699
169 2.74861645371893
170 2.24702878023091
171 1.56610952548402
172 0.733560323504616
173 -0.202483488264905
174 -1.17763828231205
175 -2.11790010717667
176 -2.94789036492338
177 -3.59964590234766
178 -4.02077866200788
179 -4.18089178595179
180 -4.07535725467226
181 -3.72590608478887
182 -3.17790960462623
183 -2.49468186444161
184 -1.74954712077158
185 -1.0167340751119
186 -0.362332238223189
187 0.163454660422632
188 0.531703801586966
189 0.737101166394163
190 0.796934712153964
191 0.746975219046367
192 0.634922098142415
193 0.512439523672607
194 0.426994816923434
195 0.414708457478572
196 0.49523767257802
197 0.669377634129153
198 0.91963302652346
199 1.21355763140512
};
\addplot [semithick, forestgreen4416044]
table {%
0 1.27722900450867
1 1.41551608097989
2 1.55356066350173
3 1.68854489844341
4 1.81806686953129
5 1.94021806348675
6 2.05340615477955
7 2.15644439928135
8 2.24839944347343
9 2.32865925584902
10 2.39681989355948
11 2.45271575997486
12 2.49636163855956
13 2.52793569058151
14 2.54778117429524
15 2.55633680032274
16 2.5541895548658
17 2.54195621405449
18 2.52037229979484
19 2.4901425257407
20 2.45205043325116
21 2.40680756211834
22 2.35516915426353
23 2.29781691350973
24 2.23546754060901
25 2.16882191764263
26 2.09865553417841
27 2.02585651664294
28 1.9514873824154
29 1.87691167609637
30 1.80380923367669
31 1.73434779028383
32 1.6711042718581
33 1.6171578477341
34 1.57581062686065
35 1.55041100326962
36 1.54382238746529
37 1.55801298877834
38 1.59365285701025
39 1.64998057178866
40 1.72502303231044
41 1.81587586953721
42 1.91923620493865
43 2.03163229362321
44 2.14979825468065
45 2.27069831723403
46 2.39169549290535
47 2.51050361766872
48 2.6252310727708
49 2.7343647168067
50 2.83673225748479
51 2.93154335504429
52 3.01828655606759
53 3.09680851477926
54 3.16715683416125
55 3.22966272303206
56 3.28474820268652
57 3.33298314079851
58 3.37488948187692
59 3.41096436957848
60 3.44152808250816
61 3.4667258036838
62 3.48646388969996
63 3.50040540778045
64 3.50801317863733
65 3.5085390503055
66 3.5011564425515
67 3.48492967753744
68 3.45899559526423
69 3.42250202696064
70 3.37479557730545
71 3.31533368458455
72 3.2438449812403
73 3.16023360080911
74 3.06468989681464
75 2.95761199489449
76 2.83965660187306
77 2.71170222198449
78 2.57484284476697
79 2.43041169132242
80 2.27993344891172
81 2.12521956519422
82 1.9683062504364
83 1.81162573107042
84 1.65796402332475
85 1.51069988610311
86 1.37378321728776
87 1.25192859005796
88 1.15036235272218
89 1.07441700213944
90 1.02831874718004
91 1.01385707414917
92 1.02934591935412
93 1.07007114514707
94 1.12964969398477
95 1.2016345165291
96 1.28038262822694
97 1.36139329883942
98 1.44121924201245
99 1.51727284043345
100 1.58764653638314
101 1.6508906610756
102 1.70595287544494
103 1.7519794881726
104 1.78835268748929
105 1.81450677013751
106 1.83001789638038
107 1.83444542395525
108 1.82743227869698
109 1.80859394663403
110 1.77759304971324
111 1.7340960119155
112 1.67779634150807
113 1.60844759362188
114 1.52582928156169
115 1.42985242249744
116 1.32048871730296
117 1.19794483167686
118 1.06262395918809
119 0.915441199281808
120 0.758130713440752
121 0.594556243217596
122 0.434657102292261
123 0.30984458018008
124 0.29897537144254
125 0.427737134577846
126 0.623827806846094
127 0.847617968136122
128 1.08560876518141
129 1.33189241850115
130 1.58276121300296
131 1.83531999125636
132 2.08690131227299
133 2.33495025174234
134 2.57692596100242
135 2.8102973268009
136 3.03258422881746
137 3.24133073919299
138 3.43422414437666
139 3.60902892573531
140 3.76374732345343
141 3.89652733219723
142 4.00583013894442
143 4.09033846542848
144 4.14910046764935
145 4.1814683769489
146 4.18719835860417
147 4.16644527197122
148 4.11981262746773
149 4.04842147687017
150 3.95392286971051
151 3.8386514974444
152 3.70562901545911
153 3.5588015929755
154 3.40305670104616
155 3.24449879189846
156 3.09041426598486
157 2.94936497576815
158 2.83074480198627
159 2.74412042521973
160 2.69782622639786
161 2.69744616296308
162 2.74441172803284
163 2.83569000907329
164 2.9644952318381
165 3.12177929929148
166 3.29768596493557
167 3.48261510303136
168 3.66782887136707
169 3.84564035427211
170 4.00950674451409
171 4.1538854124711
172 4.27425060560337
173 4.36689267601689
174 4.42897541736128
175 4.45834703302948
176 4.45362588882477
177 4.4140410278964
178 4.33950967825992
179 4.23052185907895
180 4.08817964721439
181 3.91414039619604
182 3.710603285889
183 3.48032184181663
184 3.22654574591598
185 2.95311610397608
186 2.66441282960021
187 2.36559141693469
188 2.06270822216881
189 1.76337915826836
190 1.4777108922453
191 1.22045553342049
192 1.01393065254513
193 0.88856697531394
194 0.868638474181703
195 0.946348329692872
196 1.08582333105783
197 1.25192270419884
198 1.42163543940003
199 1.58156711819799
};
\end{axis}

\end{tikzpicture} &
                    \intercolarrow
                    &
                    \providecommand{\figheight}{0.9in}%
                    \providecommand{\figwidth}{2.2in}%
                    \input{\tikzdir/e2v2t_schematic/car}
                    & \\
                    ECoG recording && notch filter && high-$\gamma$ bandpass && analytic amplitude && CAR
                \end{tabular}%
            };
            \node[inner sep=0pt,draw=black,dashed,rounded corners,fit=(preprocessing)] (preprocrect) {};
            
            \node[rotate=90, below=0.82in of preprocrect.south west, minimum height=0.5cm] (w2v) {\input{\tikzdir/e2v2t_schematic/w2v}};
            \node[inner sep=0pt,draw=black,dashed,rounded corners,rotate fit=90,fit=(w2v)] (w2vrect) {};
            \node[below=0in of w2v.west] (w2vlabel) {\wavtovec};
            
            \node[below=0.5in of preprocrect.south east,anchor=north east] (downsampled) {%
                \newcommand{\figwidth}{1.7in}%
                \input{\tikzdir/e2v2t_schematic/downsample}%
            };
            \node[below=0 in of downsampled, text width=6cm, align=center] (down) {\small downsample $\rightarrow$ 100 Hz};

            \node[right=0.15in of w2vrect.south] (e2tA)
            {\begin{tikzpicture}
    
    \node[rotate=180, trapezium, minimum height=0.5cm, minimum width=2cm, draw, fill=blue!20] (e2tencoder) {\rotatebox{180}{encoder}};
    \node[below=0.2in of e2tencoder, trapezium, minimum height=0.5cm, minimum width=2cm, draw, fill=green!20, yshift=-0.5cm] (e2tdecoder) {{decoder}};
    
    \node[below=0.15in of e2tdecoder] (e2toutput) {\small `why', `yell', ...};
    
    \draw[->] (e2tencoder) -- (e2tdecoder);
    \draw[->] (e2tdecoder) -- (e2toutput);
\end{tikzpicture}};
            \node[inner sep=0pt,draw=black,dashed,rounded corners,fit=(e2tA)] (e2tArect) {};
            \node[below=0in of e2tArect.south] (e2tAlabel) {\ecogtotxt};

            \node[left=0.15in of downsampled.west] (e2tB)
            {\begin{tikzpicture}
    
    \node[rotate=180, trapezium, minimum height=0.5cm, minimum width=2cm, draw, fill=blue!20] (e2tencoder) {\rotatebox{180}{encoder}};
    \node[below=0.2in of e2tencoder, trapezium, minimum height=0.5cm, minimum width=2cm, draw, fill=green!20, yshift=-0.5cm] (e2tdecoder) {{decoder}};
    
    \node[below=0.15in of e2tdecoder] (e2toutput) {\small `why', `yell', ...};
    
    \draw[->] (e2tencoder) -- (e2tdecoder);
    \draw[->] (e2tdecoder) -- (e2toutput);
\end{tikzpicture}};
            \node[inner sep=0pt,draw=black,dashed,rounded corners,fit=(e2tB)] (e2tBrect) {};
            \node[below=0in of e2tBrect.south] (e2tBlabel) {\ecogtotxt};

            \draw ($(e2tArect.north east)!.5!(e2tBrect.north west)$) --($(e2tArect.south east)!.5!(e2tBrect.south west)$);
            
            \node[below left=0.05 and 0.1in of preprocrect.south west] (fakeA) {};
            \node[below right=0.05in and 0.1in of preprocrect.south east] (fakeB) {};
            \draw[->] (preprocrect.east) -| (fakeB.center);
            \draw[->] (fakeB.center) |- (fakeA.center) |- (w2vrect.north);
            \draw[->] (fakeB.center) |- (downsampled.east); 

            \node[above left=0.1in and 0.02in of e2tArect.north west] (fakeC) {};
            \node[above=0.15in of e2tArect.north] (fakeD) {};
            \draw[->] (w2vrect.south) -| (fakeC.center) |- (fakeD.center) -| (e2tArect.north); 

            \node[above right=0.1in and 0.02in of e2tBrect.north east] (fakeE) {};
            \node[above=0.15in of e2tBrect.north] (fakeF) {};
            \draw[->] (downsampled.west) -| (fakeE.center) |- (fakeF.center) -| (e2tBrect.north);

        \end{tikzpicture}
        \caption{Training pipelines.}
        \label{fig:schematic}
    \end{figure}

\subsection{Evaluation}
Downstream performance of the learned representations was evaluated under the same encoder-decoder model used in \cite{Makin2020}, which for brevity we refer to henceforth as \ecogtotxt, with a few minor alterations.
In that study, the encoder and decoder consisted of LSTM-based recurrent neural networks, coupled by the final hidden state of the encoder.
The encoder was trained to predict a representation of the speech audio (the mel-frequency cepstral coefficients, MFCCs) while the decoder was trained to predict the next word in autoregressive fashion.

The alterations are as follows.
Health privacy laws preclude the sharing of speech audio, so we substituted the time-aligned phoneme transcriptions for the MFCCs.
This caused a small decrement in performance for \partc.
We also replaced the LSTMs with GRUs and coupled the encoder and decoder with cross attention rather than final hidden state.
These improved performance very mildly.
We train for 400 epochs instead of 800, but found this has no effect on WERs.
Finally, to match \wavtovec\ context vectors (see above), all input (neural) data were sampled at $\sim$100 Hz rather than 200 Hz as in the original study; this had a deleterious effect on some subjects, especially \parta.
The original implementation \cite{Makin2020} was in {\sc TensorFlow}; our implementation is in {\sc PyTorch}.

\paragraph{Control.}
To determine whether \wavtovec's context vectors provide a superior representation to the original (preprocessed) ECoG data, we trained \ecogtotxt\ models on both.
In particular, we took preprocessed ECoG data (see \sctn{data} above) and downsampled it to 100 Hz, matching the \wavtovec\ representations, and then trained otherwise identical models on one or the other of these two types of neural data.

\paragraph{Varying training-set sizes for \ecogtotxt.}
The total number of data collected from each participant is arbitrary in that it was determined by the vagaries of their hospital stays, and varied from participant to participant.
We are primarily interested in decoding performance, then, under all possible sizes, rather than merely the maximal size, of each training set.
Consequently, we train \ecogtotxt\ models on subsets of various sizes of the in-distribution blocks, spanning the range from very few blocks to all blocks (excluding the validation blocks).
We call each choice of number of training blocks an ``experiment.''

\paragraph{Cross validation.}
For all ``experiments,'' we trained and cross-validated 30 separate \ecogtotxt\ models from scratch on the \wavtovec\ context vectors; and another 30 \ecogtotxt\ models on the preprocessed ECoG data (as in the original study) as a control.
For example, participant \parta\ completed just three blocks of the in-distribution data (MOCHA-1), so we conducted just two experiments.
In the first, for each possible choice of validation block, we trained 10$\times$2 models on one of the two remaining blocks (randomly chosen), yielding 30$\times$2 models.
In the second, for each possible choice of validation block, we trained 10$\times$2 models on both of the two remaining blocks, yielding another 30$\times$2 models.
Participant \parta\ completed ten blocks of the in-distribution data (MOCHA-1) and accordingly yielded 10 experiments.

In-distribution blocks for participants \partc\ and \partd\ consisted of either all 30 picture-description sentences or a subset of just 10.
Following the original study, we chose the validation blocks exclusively from the latter.
These subjects had far more training blocks; we ran eight experiments for each, spanning the range from just 4, up to all, training blocks.

\paragraph{Hardware.}
All \wavtovec\ models were trained on a {\sc{Nvidia RTX A6000}} GPU.
Each model required roughly 2--3 hours to train to convergence, and used around 10 gigabytes of memory.

Each \ecogtotxt\ model was cross-validated on a single V100 GPU, typically in 5--10 minutes of wall time.

\section{Results}\label{sec:results}

\subsection{Single-subject decoding}
\begin{figure}[ht!]
        \centering
        \newcommand{\figwidth}{\linewidth}
        \newcommand{\figheight}{2.5in}
\begin{tikzpicture}
\centering

\definecolor{gray}{RGB}{128,128,128}
\definecolor{green01270}{RGB}{0,127,0}
\definecolor{lightgray204}{RGB}{204,204,204}
\definecolor{orange}{RGB}{255,165,0}
\providecommand{\figheight}{7cm}
\providecommand{\figwidth}{12cm}

\begin{axis}[
legend cell align={left},
legend style={fill opacity=0.8, draw opacity=1, text opacity=1, draw=lightgray204},
tick align=outside,
tick pos=left,
x grid style={gray},
xlabel={training size (no.\ of blocks)},
xmajorgrids,
xmin=-1.05, xmax=44.05,
xtick style={color=black},
y grid style={gray},
ylabel={WER},
ymajorgrids,
ymin=0, ymax=1,
ytick style={color=black},
width=\figwidth,
height=\figheight
]
\addplot [semithick, orange, mark=+, mark size=3, mark options={solid, dashed}]
table {%
1 0.800214309692383
2 0.677091331481934
};
\addplot [semithick, orange, mark=asterisk, mark size=3, mark options={solid}]
table {%
1 0.784813537597656
2 0.691269874572754
};
\addplot [semithick, blue, mark=+, mark size=3, mark options={solid}, dashed]
table {%
1 0.647821464538574
2 0.48514289855957
3 0.323190479278564
4 0.23975001335144
5 0.142706346511841
6 0.11090078830719
7 0.0734895396232605
8 0.0545674610137939
9 0.0520595240592957
};
\addplot [semithick, blue, mark=asterisk, mark size=3, mark options={solid}]
table {%
1 0.49023811340332
2 0.252980146408081
3 0.134559526443481
4 0.0931031799316406
5 0.072503969669342
6 0.0528744626045227
7 0.0572698450088501
8 0.0349206352233887
9 0.0330966830253601
};
\addplot [semithick, green01270, mark=+, mark size=3, mark options={solid}, dashed]
table {%
4 0.825794563293457
7 0.648777351379395
10 0.513231626698669
13 0.390813985879971
16 0.324130296438513
19 0.277699045722867
22 0.253126345889669
25 0.230939197540283
};
\addplot [semithick, green01270, mark=asterisk, mark size=3, mark options={solid}]
table {%
4 0.764737288571714
7 0.487518921192725
10 0.3529259568169
13 0.25304879023705
16 0.2363224572419
19 0.204095826031249
22 0.153328975571526
25 0.138403532874416
};
\addplot [semithick, purple, mark=+, mark size=3, mark options={solid}, dashed]
table {%
4 0.792646046094997
9 0.693141805274146
15 0.520194394247872
20 0.358757600565066
26 0.265628128051758
31 0.21561488310496
37 0.170093234380086
42 0.12895401318868
};
\addplot [semithick, purple, mark=asterisk, mark size=3, mark options={solid}]
table {%
4 0.751929437546503
9 0.534340088786359
15 0.38696913421154
20 0.315420649517542
26 0.250975282986959
31 0.214348390201728
37 0.170308642727988
42 0.16116935752687
};

\end{axis}
\node[draw, fill=white, inner sep=5pt] at (11,2.8) {%
    \begin{minipage}{0.7in}
        \begin{tikzpicture}
          \draw[semithick, black, dashed] (0.5,-0.5) -- ++(0.5,0) node[right, font=\footnotesize] {ECoG};
          \draw[semithick, black] (0.5,-0.9) -- ++(0.5,0) node[right, font=\footnotesize] {context};
          \fill[orange] (0.5,-1.3) circle [xshift=0.25cm, radius=2pt] node[xshift=0.475cm, right, font=\footnotesize] {\textcolor{black}{\parta}};
          \fill[blue] (0.5,-1.7) circle [xshift=0.25cm, radius=2pt] node[xshift=0.475cm, right, font=\footnotesize] {\textcolor{black}{\partb}};
          \fill[green01270] (0.5,-2.1) circle [xshift=0.25cm, radius=2pt] node[xshift=0.475cm, right, font=\footnotesize] {\textcolor{black}{\partc}};
          \fill[purple] (0.5,-2.5) circle [xshift=0.25cm, radius=2pt] node[xshift=0.475cm, right, font=\footnotesize] {\textcolor{black}{\partd}};
        \end{tikzpicture}
    \end{minipage}
};

\end{tikzpicture}
        \caption{Median decoding performance (word error rate [WER]) for each participant using preprocessed ECoG (dashed) and \wavtovec\ context vectors (solid), as a function of number of training blocks.
        Study participants are indicated by color.
        }
        \label{fig:performanceVsTrainingSize}
    \end{figure}
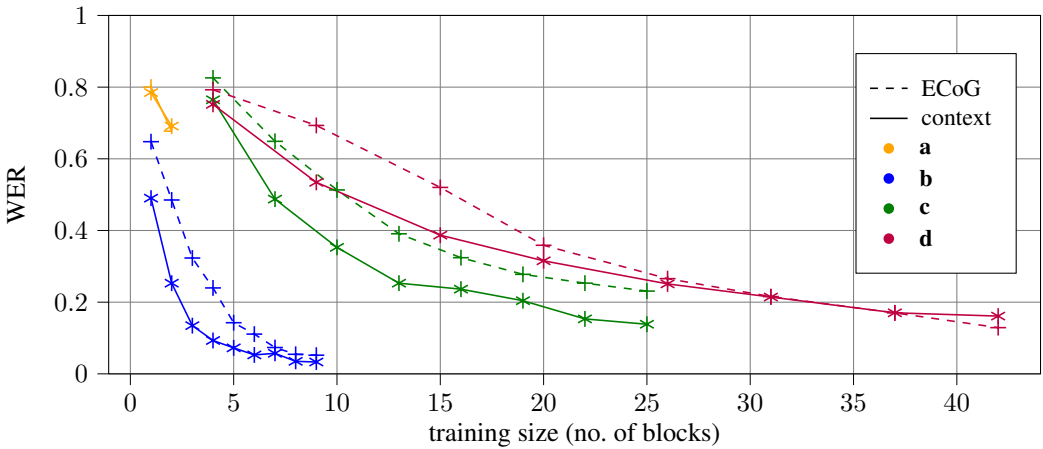

We begin by summarizing results across all participants and experiments.
\fig{performanceVsTrainingSize} shows word error rates (WERs) for \ecogtotxt\ as a function of the number of blocks used for (supervised) training.
The solid lines reproduce the results of \cite{Makin2020}, with some small discrepancies due to the small changes in our preprocessing and implementation of \ecogtotxt\ (see \sctn{methods}).
Each star corresponds to the median over 30 different instances of \ecogtotxt\ and choices for the validation block(s).
The dashed lines show the corresponding results when training on the context vectors produced by \wavtovec\ instead of the original ECoG data.
For nearly all choices of training size across all participants, context vectors yield better decoding, and for no experiments do the original ECoG data produce better performance than the context vectors.
The median improvements in WER (as a percentage of the WER for ECoG data) are given in \tbl{performanceVsTrainingSize}, along with significance of the difference in WER distributions (Wilcoxon signed-rank test).

We consider participants individually (see also \fig{cv_t08_p50} in the appendix).
For particular \partb, for all training sizes except one, the context vectors yielded  significantly better decoding than the ECoG data. 
With a training size of 2 blocks, the median improvement is $\sim$45\%, bringing word error rates down near the outer bound of acceptable speech transcription \cite{Munteanu06}.
This is a larger improvement than even that provided by cross-participant transfer learning in the original study \cite{Makin2020}.
We note that this participant also had the fewest OOD data for training \wavtovec\ (half an hour compared to the other participants' hour).
Participant \partc\ saw significantly better performance at every training size.
In contrast, no transfer learning scheme presented in the original study \cite{Makin2020} improved decoding results for this participant.
For participant \partd, as for \partb, improvements are significant with fewer (supervised-)training blocks, but disappear as more are used.
This suggests that for these participants, performance on this circumscribed decoding task has saturated at (respectively) about 10\% and 2\% word error rates.
We return to this point in the discussion.
This does not generically explain the failures of the context vectors to improve performance, however:
For participant \parta, WERs with context vectors and the original ECoG data are not distinguishable, yet we know much lower WERs are achievable on this task from these data \cite{Makin2020}.

\TablePerformanceVsTrainingSize

\paragraph{The usefulness of OOD labels.}
The blocks that we designate as OOD do, in fact, have labels.
We have thrown them away to simulate a scenario in which labels do not exist, but now we ask how much these labels help if we avail ourselves of them.
In particular, we compare using the OOD data without labels to train \wavtovec\, as we have been doing until this point; vs.\ using them with labels as additional training blocks for \ecogtotxt\ (\ie\ along with the MOCHA-1 blocks used heretofore).
We focus on MOCHA-TIMIT and furthermore use only two blocks of MOCHA-1 in supervised training in order to reprise the comparisons made in Fig.\ 3 of the original study.
Here, as there, we are interested in how performance can be improved when the floor has not been reached.
Note that we include \partd\ in these MOCHA-TIMIT results:\ although the large majority of this participant's blocks were picture descriptions, three were MOCHA-TIMIT blocks.

    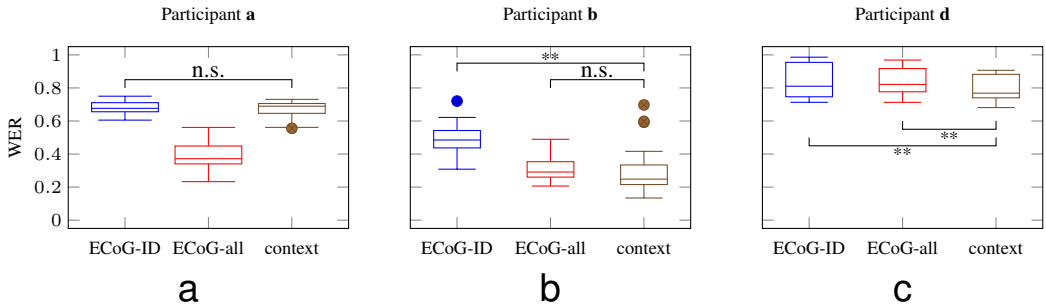
\begin{figure}[h!]
    \centering
    \providecommand{\figwidth}{0.38\linewidth}%
    \providecommand{\figheight}{4cm}%
    \subfloat[]{%
        \scriptsize%
        \label{subfig:mochaparta}%
        \begin{tikzpicture}

    \providecommand{\figwidth}{4cm}
    \providecommand{\figheight}{5cm}
    \begin{axis}[
        boxplot/draw direction=y,
        width=\figwidth,
        height=\figheight,
        title={Participant \parta},
        ytick={0, 0.2,0.4,0.6,0.8,1.0},
        xtick={1,2,3},
        ymin=-0.05, ymax=1.05,
        xticklabel style = {align=center,text height=0.1in, text width=0.8in,minimum width=0.8in},
        xticklabels={ECoG-ID, ECoG-all, context},
        ylabel={WER},
        ylabel near ticks,
        ]

        \addplot+ [boxplot]
        table [row sep=\\, y index=0] {
            0.7414524841308594\\ 0.6701825714111328\\ 0.6915079498291016\\ 0.7108809661865234\\ 0.6551428985595703\\ 0.7500477600097656\\ 0.712626953125\\ 0.6840555572509766\\ 0.7109207153320313\\ 0.646547622680664\\ 0.6616032409667969\\ 0.6766746520996094\\ 0.6877778625488281\\ 0.6695635986328125\\ 0.693198471069336\\ 0.6775080108642578\\ 0.7263096618652344\\ 0.6821190643310547\\ 0.717293701171875\\ 0.7114445495605469\\ 0.6623651123046875\\ 0.7179127502441406\\ 0.6552619171142579\\ 0.605285758972168\\ 0.6686746215820313\\ 0.6646745300292969\\ 0.611595230102539\\ 0.6565158081054687\\ 0.6543651580810547\\ 0.6298888778686523\\
        };

        \addplot+ [boxplot]
        table [row sep=\\, y index=0] {
            0.443039665222168\\ 0.45446029663085935\\ 0.41289684295654294\\ 0.46269046783447265\\ 0.4726348876953125\\ 0.5243968200683594\\ 0.5167382049560547\\ 0.5612380981445313\\ 0.4812697982788086\\ 0.47361907958984373\\ 0.2937460708618164\\ 0.23307937622070313\\ 0.33586505889892576\\ 0.35755558013916017\\ 0.3321666717529297\\ 0.30885713577270507\\ 0.35363494873046875\\ 0.39514286041259766\\ 0.32901588439941404\\ 0.3075476264953613\\ 0.3721111297607422\\ 0.3720396423339844\\ 0.3682618713378906\\ 0.3661745834350586\\ 0.34825389862060546\\ 0.40707141876220704\\ 0.43301589965820314\\ 0.39903968811035156\\ 0.3653888702392578\\ 0.3450873184204102\\
        };

        \addplot+ [boxplot]
        table [row sep=\\, y index=0] {
            0.6629206848144531\\ 0.6680715942382812\\ 0.6639206695556641\\ 0.721198501586914\\ 0.716031723022461\\ 0.7256825256347657\\ 0.6949444580078125\\ 0.7114920806884766\\ 0.6568889617919922\\ 0.6463809967041015\\ 0.6695793914794922\\ 0.5946429443359375\\ 0.5620158767700195\\ 0.5613809967041016\\ 0.630682601928711\\ 0.6896349334716797\\ 0.6718968963623047\\ 0.6462064361572266\\ 0.555666732788086\\ 0.6401111602783203\\ 0.7000476837158203\\ 0.6933811187744141\\ 0.7015953826904296\\ 0.7186191558837891\\ 0.7029048919677734\\ 0.730999984741211\\ 0.6929048156738281\\ 0.7086190032958984\\ 0.7086190795898437\\ 0.6962381744384766\\
        };

    \draw [] (axis cs:1,0.8) -- (axis cs:1,0.85) -- (axis cs:3,0.85) -- (axis cs:3,0.8);
    \node at (axis cs:2,0.9) {\small n.s.};


    \end{axis}
\end{tikzpicture}%
%
    }%
    \subfloat[]{%
        \label{subfig:mochapartb}%
        \scriptsize%
        \pgfplotsset{every axis post/.append style={yticklabels = {} }}%
        \begin{tikzpicture}%

    \providecommand{\figwidth}{4cm}%
    \providecommand{\figheight}{5cm}%
    \begin{axis}[
        boxplot/draw direction=y,
        boxplot/box extend=0.5,
        title={Participant \partb},
        ytick={0, 0.2,0.4,0.6,0.8,1.0},
        ymin=-0.05, ymax=1.05,
        xtick={1,2,3},
        xticklabel style = {align=center,text height=0.1in, text width=0.8in,minimum width=0.8in},
        xticklabels={ECoG-ID, ECoG-all, context},
        width=\figwidth,
        height=\figheight,
    ]
    
    \addplot+ [boxplot] table [row sep=\\, y index=0] {
        0.6214762496948242\\ 0.5918174743652344\\ 0.7205477142333985\\ 0.48818038940429687\\ 0.49762989044189454\\ 0.4095584487915039\\ 0.30833332061767577\\ 0.4733651351928711\\ 0.5088810348510742\\ 0.5687539672851563\\ 0.5394207000732422\\ 0.45838096618652346\\ 0.6073889541625976\\ 0.42395236968994143\\ 0.5331349563598633\\ 0.485174674987793\\ 0.41584918975830076\\ 0.4621587753295898\\ 0.43382537841796875\\ 0.4637460708618164\\ 0.397769889831543\\ 0.48511112213134766\\ 0.5454206466674805\\ 0.5055397796630859\\ 0.417309455871582\\ 0.6073096084594727\\ 0.44050796508789064\\ 0.576912727355957\\ 0.4482777786254883\\ 0.4842619323730469\\
    };
    
    \addplot+ [boxplot] table [row sep=\\, y index=0] {
        0.4814682769775391\\ 0.41665866851806643\\ 0.4891746139526367\\ 0.20637374877929687\\ 0.26788169860839844\\ 0.2686356163024902\\ 0.38013496398925783\\ 0.3490476226806641\\ 0.36156352996826174\\ 0.26848411560058594\\ 0.2953730010986328\\ 0.257261905670166\\ 0.2711666679382324\\ 0.30448410034179685\\ 0.2901269721984863\\ 0.245317440032959\\ 0.23838886260986328\\ 0.21211111068725585\\ 0.3038333511352539\\ 0.22711111068725587\\ 0.2635079574584961\\ 0.3586190414428711\\ 0.32599998474121095\\ 0.36546031951904295\\ 0.29134923934936524\\ 0.37346031188964846\\ 0.3298094940185547\\ 0.2666587257385254\\ 0.3338888549804688\\ 0.22667461395263672\\
    };

    \addplot+ [boxplot] table [row sep=\\, y index=0] {
        0.592468376159668\\ 0.5978571701049805\\ 0.6971190643310546\\ 0.22060392379760743\\ 0.2581031608581543\\ 0.29488090515136717\\ 0.2330238151550293\\ 0.3492142868041992\\ 0.31018251419067383\\ 0.18908729553222656\\ 0.17951587677001954\\ 0.24372220993041993\\ 0.22887302398681642\\ 0.2810873031616211\\ 0.18848413467407227\\ 0.21117462158203126\\ 0.21962696075439453\\ 0.1811111068725586\\ 0.3455237579345703\\ 0.37496826171875\\ 0.26666669845581054\\ 0.13397619247436524\\ 0.3261745834350586\\ 0.2334285545349121\\ 0.3414841461181641\\ 0.24119049072265625\\ 0.24834918975830078\\ 0.41632537841796874\\ 0.17869842529296875\\ 0.25761110305786133\\
    };

    \draw [] (axis cs:1,0.9) -- (axis cs:1,0.95) -- (axis cs:3,0.95) -- (axis cs:3,0.9);
    \node at (axis cs:2,0.975) {**};

    \draw [] (axis cs:2,0.8) -- (axis cs:2,0.85) -- (axis cs:3,0.85) -- (axis cs:3,0.8);
    \node at (axis cs:2.5,0.9) {\small n.s.};

    \end{axis}
\end{tikzpicture}%
%
    }%
    \subfloat[]{%
        \label{subfig:mochapartd}%
        \scriptsize%
        \pgfplotsset{every axis post/.append style={yticklabels = {} }}%
        \begin{tikzpicture}%
    \providecommand{\figwidth}{4cm}%
    \providecommand{\figheight}{5cm}%
    \begin{axis}[
        boxplot/draw direction=y,
        boxplot/box extend=0.5,
        ytick={0, 0.2,0.4,0.6,0.8,1.0},
        title={Participant \partd},
        ymin=-0.05, ymax=1.05,
        xtick={1,2,3},
        xticklabel style = {align=center,text height=0.1in, text width=0.8in,minimum width=0.8in},
        xticklabels={ECoG-ID, ECoG-all, context},
        width=\figwidth,
        height=\figheight,
    ]
    \addplot+ [boxplot] table [row sep=\\, y index=0] {
        0.9597221374511719\\ 0.9430793762207031\\ 0.9757936859130859\\ 0.9642301177978516\\ 0.949730224609375\\ 0.9657936859130859\\ 0.9862541198730469\\ 0.96984130859375\\ 0.9636190032958984\\ 0.9729367065429687\\ 0.8105794525146485\\ 0.7900794982910156\\ 0.7794921112060547\\ 0.7980317687988281\\ 0.8115160369873047\\ 0.8513492584228516\\ 0.7786111450195312\\ 0.824635009765625\\ 0.8309049224853515\\ 0.8655556488037109\\ 0.7274633026123047\\ 0.7130592346191407\\ 0.7184236145019531\\ 0.7428600311279296\\ 0.7458759307861328\\ 0.7378998565673828\\ 0.7502806854248046\\ 0.773177490234375\\ 0.7473045349121094\\ 0.7374632263183594\\
    };
    \addplot+ [boxplot] table [row sep=\\, y index=0] {
        0.9153970336914062\\ 0.9386270904541015\\ 0.9179842376708984\\ 0.9334763336181641\\ 0.9687698364257813\\ 0.9169205474853516\\ 0.946230239868164\\ 0.9577142333984375\\ 0.9431032562255859\\ 0.9257540893554688\\ 0.7291825866699219\\ 0.7465714263916016\\ 0.7433254241943359\\ 0.7550080108642578\\ 0.8043571472167969\\ 0.8041587829589844\\ 0.8285001373291015\\ 0.7130476379394531\\ 0.7517778015136719\\ 0.8432380676269531\\ 0.7878007507324218\\ 0.8265231323242187\\ 0.8442294311523437\\ 0.8402294921875\\ 0.821348648071289\\ 0.8040072631835937\\ 0.8045866394042969\\ 0.813713607788086\\ 0.8050865173339844\\ 0.7659199523925782\\
    };
    \addplot+ [boxplot] table [row sep=\\, y index=0] {
        0.9041668701171875\\ 0.8973890686035156\\ 0.8866033172607422\\ 0.9029447174072266\\ 0.8773890686035156\\ 0.906722412109375\\ 0.8833890533447266\\ 0.8810081481933594\\ 0.8950556945800782\\ 0.9048493957519531\\ 0.750158920288086\\ 0.8148969268798828\\ 0.7961032867431641\\ 0.7698175048828125\\ 0.8288491821289062\\ 0.8016587829589844\\ 0.7457620239257813\\ 0.7613968658447265\\ 0.7616588592529296\\ 0.763730239868164\\ 0.6811898803710937\\ 0.7353645324707031\\ 0.6841580963134766\\ 0.708523178100586\\ 0.7118311309814453\\ 0.7684344482421875\\ 0.745308837890625\\ 0.7447676849365235\\ 0.7186817932128906\\ 0.707664566040039\\
    };
    
    \draw [] (axis cs:1,0.5) -- (axis cs:1,0.45) -- (axis cs:3,0.45) -- (axis cs:3,0.5);
    \node at (axis cs:2,0.4) {**};

    \draw [] (axis cs:2,0.6) -- (axis cs:2,0.55) -- (axis cs:3,0.55) -- (axis cs:3,0.6);
    \node at (axis cs:2.5,0.5) {**};

    \end{axis}%
\end{tikzpicture}%
%
    }%
    \caption{%
        The effect of OOD data on MOCHA-1 decoding performance.
        \textcolor{blue}{ECoG-ID}:\ supervised training on ``in-distribution'' data, MOCHA-1; \textcolor{red}{ECoG-all}:\ supervised training on in- as well as out-out-distribution data; \textcolor{brown}{context}: self-supervised training on OOD data, supervised training on MOCHA-1.
        All models were tested on one randomly held out MOCHA-1 block.
        Stars indicate context vectors are significantly better than other decoding schemes (Wilcoxon signed-rank test; * $p<0.05$; ** $p<0.005$; Holm-Bonferroni corrected for multiple comparisons).
        Boxes cover interquartile distances across 30 trained \ecogtotxt\ instances; whiskers extend to the last datum within 150\% of the interquartile range from the end of the box.
        Exact p-values can be found in \tbl{cross_validation_p_mocha}.
    }\label{fig:MOCHAtwoBlocks}
    \end{figure}

We have already seen that the context vectors provide no improvement over ECoG for \parta.
Using the OOD blocks for supervised training does (\subfig{mochaparta}).
However, for \partb, the large improvement provided by training on labeled OOD blocks is matched by our label-free training (\subfig{mochapartb}).
Finally, for participant \partd, \wavtovec\ context vectors, the effect size is quite small.
Nevertheless, label-free training provides better performance than doing without these blocks altogether but also better than using them with labels for supervised training (\subfig{mochapartd}).

\subsection{Transfer learning}
Since the context vectors generated for participant \parta\ did not improve decoding, we consider ways to improve them.
In particular, we attempt a form of transfer learning with \wavtovec.
The original study showed that ``pre-training'' \ecogtotxt\ on participant \partb's data before training it on participant \parta's data improves decoding performance on participant \parta.
Here we attempt an analogous transfer.
We take a \wavtovec\ model trained on participant \partb's OOD blocks, replace its initial convolutional layer with an appropriately sized random layer, and train the resulting model on \parta's OOD blocks (see \sctn{methods} for details).
We use the resulting \wavtovec\ model to transform the in-distribution MOCHA-1 blocks into context vectors, and train \ecogtotxt\ models on them, as before.
We find that when using one block for training and one for testing, WER drops ($p<0.01$) by 6.26\% from an average WER of 0.81 to an average of 0.76.
For two-block (the largest) training sets, these context vectors improve WER over the original ECoG ($p<0.005$) by 14.6\%, from an average WER of 0.68 to an average of 0.58.

However, no other combination of transfer-learning approach resulted in context vectors that significantly improved decoding performance.

\subsection{Negative results}
\TableWVlossVsETloss
It would be be useful and intuitive if lower \wavtovec\ losses presage lower word error rates under \ecogtotxt.
However, this does not appear to be the case.
We have already seen from our hyperparameter selection (\tbl{effectOfLookahead}) that decreasing $s$, the number of steps that \wavtovec\ must ``predict'' ahead, monotonically decreases the \wavtovec\ loss but not decoding errors.
But this is likewise found with speech audio \cite{vandenOord2018} and is in any case expected, since the features that are easiest to predict (slowly varying content) are unlikely to be the most relevant to the downstream task (phones).
However, across participants, even at a fixed $s=8$, lower \wavtovec\ losses do not correspond to lower WERs.
\tbl{WVlossVsETloss} shows that participant \parta, for whom WERs are both highest and least improved under \wavtovec's context vectors, has the lowest \wavtovec\ loss.
This might suggest an \emph{inverse} relationship between \wavtovec\ and \ecogtotxt\ losses, perhaps on the rationale that low \wavtovec\ losses indicate prediction of easy, but irrelevant features; but such a relationship is certainly not obvious from \tbl{WVlossVsETloss} and \fig{performanceVsTrainingSize}, and we are reluctant to try to construct one.

We also found that \wavtovec\ cannot replace the preprocessing steps (see top row of \fig{schematic}); it is best used as an additional one.
In our experiments, removing the common average reference degraded downstream decoding by a moderate to large amount, depending on the participant.
Removing the bandpass filter is fatal, presumably because the lower frequency bands of ECoG data have much higher amplitudes---and are therefore more easily predicted---but are much less relevant to speech.
Somewhat more surprisingly, in our experiments, using the filtered voltages rather than their envelope also proved fatal to learning useful context vectors.
Thus, we conclude that self-supervised learning works best in conjunction with established signal-processing techniques.

\section{Discussion}
The primary purpose of this investigation was to determine whether self-supervised learning from unlabeled data can reduce the burden of data-collection, ultimately from persons who have lost the ability to speak and would be the target population for a speech prosthesis.
The results are promising.
We have shown that self-supervised methods can learn, from unlabeled ECoG data recorded during speech, representations that facilitate subsequent speech decoding.
These results held for all subjects, although for one (\parta) it was necessary to learn these representations from, in part, data from a different participant (\partb).
Crucially, this also shows that transfer learning is possible in the context of self-supervised learning from ECoG data.

We note, however, a number of important limitations.
First, the speech-decoding task under consideration is highly circumscribed.
A single set of 30--50 sentences was read during both (supervised) training blocks and testing blocks.
Thus, although \ecogtotxt\ decodes word by word and from vocabularies of up to $\sim$1800 words, in practice it sometimes memorizes chunks of sentences, suggesting that it has overfit to them.
This limitation was noted in the original study \cite{Makin2020}; we emphasize it here because it likewise limits the generality of our results.
On the other hand, none of these sentences were read during the blocks that provided data for self-supervised learning.
The fact that self-supervised learning nevertheless improved performance on the 30--50 ``in-distribution'' sentences shows that it is extracting a representation that is generically useful for speech decoding.

Second, the benefits of self-supervised learning shown here are sometimes reduced or abolished when training on the full set of ``in-distribution'' data (\fig{performanceVsTrainingSize}).
We emphasize, however, that this appears to be a floor effect induced by the comparative simplicity of the task, just discussed.

Third, the ``unlabeled'' (out-of-distribution) data were in fact from controlled sessions of read speech (see \sctn{methods}).
For self-supervised learning to have its maximal impact, it must also work with unconstrained speech (or, better still, with ECoG data during other behaviors as well).
Such data are rare but we see this as perhaps the most important avenue for further research along these lines.

Finally, the participants in this study had epilepsy but were otherwise normal speakers of English.
All ECoG data were collected during overt speech (and the silences in the interstices for self-supervised learning).
In contrast, the target population will not be able to speak out loud.
Whether self-supervised learning will work as well in this context is unknown.
We are encouraged, however, by recent clinical trials in which methods developed for decoding from epilepsy patients \cite{Makin2020} were successfully ported to anarthric or dysarthric patients \cite{Moses2021,Metzger2023,Willett2023}.
Furthermore, the transfer of self-supervised learning across participants shown here suggests that data collected from epilepsy patients may be useful for training models for those who have lost the ability to speak.

\bibliographystyle{plain}
\bibliography{%
    \bibsdir/audition,%
    \bibsdir/compneuro,%
    \bibsdir/machinelearning,%
    \bibsdir/neuroscience,%
    \bibsdir/nonpapers,%
    \bibsdir/BMI,%
    \bibsdir/misc_refs
}
\newpage

\appendix

\section{Appendix}\label{sec:appendixA}
\setcounter{figure}{0}
\setcounter{table}{0}
\renewcommand{\thetable}{S\arabic{table}}
\renewcommand{\thefigure}{S\arabic{figure}}


    \begin{figure}[ht]
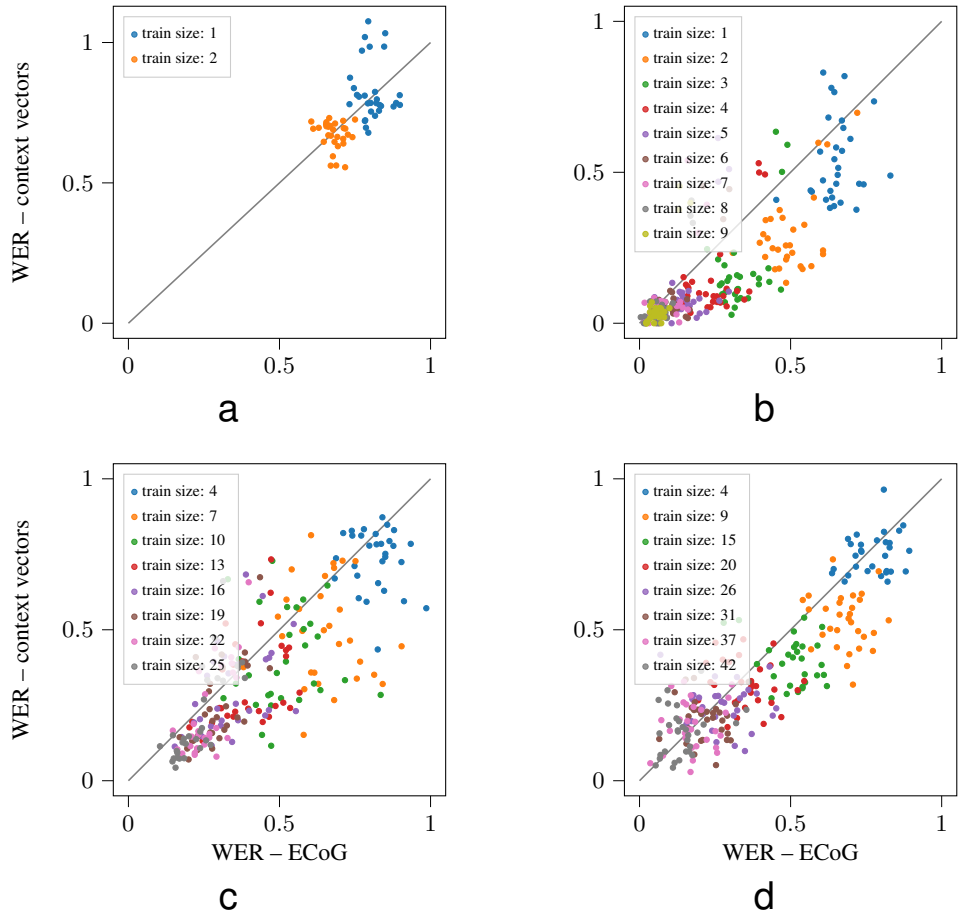

    \centering
    \footnotesize
    \subfloat[][]{
        \label{subfig:cv_car_400_t08_p50}
\begin{tikzpicture}

\definecolor{darkgray176}{RGB}{176,176,176}
\definecolor{darkorange25512714}{RGB}{255,127,14}
\definecolor{gray}{RGB}{128,128,128}
\definecolor{lightgray204}{RGB}{204,204,204}
\definecolor{steelblue31119180}{RGB}{31,119,180}
\providecommand{\figwidth}{6cm}
\providecommand{\figheight}{6cm}

\begin{axis}[
legend cell align={left},
legend style={
  fill opacity=0.8,
  draw opacity=1,
  text opacity=1,
  at={(0.03,0.97)},
  anchor=north west,
  draw=lightgray204
},
tick align=outside,
tick pos=left,
x grid style={darkgray176},
xmin=-0.05, xmax=1.05,
xtick style={color=black},
y grid style={darkgray176},
ylabel={WER -- context vectors},
ymin=-0.053772216796875, ymax=1.12921655273438,
ytick style={color=black},
width=6cm, height=6cm
]
\addplot [draw=steelblue31119180, fill=steelblue31119180, mark=*, mark size = 1, mark size = 1, only marks]
table{%
x  y
0.831460342407227 0.7775
0.792539749145508 0.783095321655273
0.822253952026367 0.7975634765625
0.886142959594727 0.784785690307617
0.763222351074219 0.807023849487305
0.898555526733398 0.777666778564453
0.756460342407227 0.813380966186523
0.877492141723633 0.772015914916992
0.897992248535156 0.812333374023438
0.836182632446289 0.773865280151367
0.849976348876953 1.03303176879883
0.782690505981445 1.01943649291992
0.772904891967773 0.971039581298828
0.782500152587891 0.720666732788086
0.816285858154297 0.738976211547852
0.794047698974609 0.678753967285156
0.793793792724609 1.0754443359375
0.783841400146484 0.722976226806641
0.786158905029297 0.696690521240234
0.846484146118164 0.985444488525391
0.733500061035156 0.874619140625
0.78207145690918 0.810793762207031
0.829404830932617 0.75696044921875
0.820603256225586 0.783984298706055
0.801880950927734 0.784841384887695
0.798547668457031 0.985126953125
0.746230239868164 0.837619171142578
0.816190567016602 0.82412712097168
0.802992095947266 0.754111175537109
0.731658782958984 0.780261993408203
};
\addlegendentry{\tiny train size: 1}
\addplot [draw=darkorange25512714, fill=darkorange25512714, mark=*, mark size = 1, only marks]
table{%
x  y
0.741452484130859 0.662920684814453
0.670182571411133 0.668071594238281
0.691507949829102 0.663920669555664
0.710880966186523 0.721198501586914
0.65514289855957 0.716031723022461
0.750047760009766 0.725682525634766
0.712626953125 0.694944458007812
0.684055557250977 0.711492080688477
0.710920715332031 0.656888961791992
0.646547622680664 0.646380996704102
0.661603240966797 0.669579391479492
0.676674652099609 0.594642944335937
0.687777862548828 0.56201587677002
0.669563598632812 0.561380996704102
0.693198471069336 0.630682601928711
0.677508010864258 0.68963493347168
0.726309661865234 0.671896896362305
0.682119064331055 0.646206436157227
0.717293701171875 0.555666732788086
0.711444549560547 0.64011116027832
0.662365112304688 0.70004768371582
0.717912750244141 0.693381118774414
0.655261917114258 0.70159538269043
0.605285758972168 0.718619155883789
0.668674621582031 0.702904891967773
0.664674530029297 0.730999984741211
0.611595230102539 0.692904815673828
0.656515808105469 0.708619003295898
0.654365158081055 0.708619079589844
0.629888877868652 0.696238174438477
};
\addlegendentry{\tiny train size: 2}
\addplot [semithick, gray, forget plot]
table {%
0 0
1 1
};
\end{axis}

\end{tikzpicture}
    }
    \hspace{0.5in}
    \subfloat[][]{
        \label{subfig:cv_car_401_t08_p50}
        \input{\tikzdir/cv_car_results_t08_p50/cv_401}
    }\\
    \subfloat[][]{
        \label{subfig:cv_car_402_t08_p50}
        \input{\tikzdir/cv_car_results_t08_p50/cv_402}
    }
    \hspace{0.5in}
    \subfloat[][]{
        \label{subfig:cv_car_403_t08_p50}
        \input{\tikzdir/cv_car_results_t08_p50/cv_403}
    }
    \caption{%
        Relative decoding performance with ECoG and \wavtovec\ context vectors.
        Each dot corresponds to an ``experiment,'' \ie\ a choice of training and validation blocks, and the corresponding word error rates for a pair of models trained on ECoG (horizontal) and context vectors (vertical).
        Panels correspond to the four study participants.
    }\label{fig:cv_t08_p50}
    \end{figure}

\TablePvals

\TableMochaPvals

\end{document}